\documentclass[intlimits,twoside,a4paper]{article}
\usepackage[cp1251]{inputenc}

\usepackage[eqsecnum]{cmpj3}

\usepackage{bm}

\issue{2018}{21}{1}{13401}
\doinumber{10.5488/CMP.21.13401}

\title[Light propagation in SPR sensor with LC]{Numerical modelling of light propagation in surface plasmon resonance sensor with liquid crystal}
\author[Ye.S. Yarmoshchuk, V.I. Zadorozhnii, V.Yu. Reshetnyak]{Ye.S. Yarmoshchuk, V.I. Zadorozhnii, V.Yu. Reshetnyak}
\address{Taras Shevchenko National University of Kyiv, Faculty of Physics, \\ 2 Acad. Glushkov Ave., 03022 Kyiv,  Ukraine}

\date{Received June 6, 2017, in final form October 4, 2017}

\begin{document}

\maketitle

\begin{abstract}
The five-layer nanorod-mediated surface plasmon sensor with inhomogeneous liquid crystal layer was theoretically investigated. The reflectance as the function of the incident angle was calculated at different voltages applied to the liquid crystal (LC) for different analyte refractive indices. By changing the LC director orientation one can control the position of the reflective dips and choose the one that is the most sensitive to the analyte refractive index. At the chosen angle of incidence, the analyte refractive index can be found from the reflectance value. The director reorientation effect is stronger when the prism refractive index is between ordinary and extraordinary refractive indices of the LC. In this case, the voltage increase and the prism refractive index decrease have a similar effect on the reflectance features.
\keywords liquid crystal, surface plasmon resonance, sensor, nanoparticles, porous metal film
\pacs 42.79.Kr, 61.30.Gd, 78.15.+e, 73.20.Mf, 42.79.Pw, 87.85.fk
\end{abstract}

\section{Introduction}

Surface plasmons (SPs) are collective electron excitations that exist at the interface between metal and dielectric. Surface plasmons play an important role in optical properties of metals, SPs are widely used in various devices, such as sensors. The attenuated total reflection (ATR) method is one of the ways to excite SPs \cite{1,2}. Widely used Kretschmann and Otto configurations are based on this method \cite{3,4,5}.
The Kretschmann configuration contains a consecutively placed coupling prism, a thin metal film and a dielectric layer (e.g., air). In the Otto configuration, a coupling prism and a metal 
layer are separated by an air gap. In both configurations, the SP is excited at the metal-dielectric (air) interface. The Otto configuration has the advantage of tunability by changing the air gap.
When the surface plasmon momentum and the tangential component of the photon momentum are equal, surface plasmon resonance occurs. As a result, there is a dip in the reflectance. The dip position is sensitive to the refractive index of the adjacent layer. Its shift can serve as the indicator of reactions that occur on the surface between the metal and the investigated medium (analyte).
This phenomenon is widely used in the development of chemical and biological sensors \cite{61,6}.
The effect of different sensor characteristics, for example metal layer parameters, the prism refractive index, on the sensor sensitivity was presented in the paper~\cite{9}.

At present, different modifications of sensors have been studied in order to enhance their sensitivity and accuracy, reduce their size.
In particular, long-range surface plasmon resonance (LRSPR) sensor with sharp reflection spectrum was suggested (see reference~\cite{9a} and reference within). This sensor contains a prism, a dielectric layer, a metal layer and an analyte in sequence.
During the last years, much attention has also been paid to localized surface plasmon (LSP) sensors that are quite promising \cite{7,8}.
The LSPs excitation in the nanostructures leads to the field enhancement near the surface. As a result, the sensor response will be stronger, with an increase of the contact surface.
Sensor with the layer of periodic gold nanowires \cite{10} presents sensitivity enhancement that depends on the period of nanowires.
Sensor with porous metal film, where pores are filled with the analyte \cite{11,12}, presents sensitivity enhancement by a factor of about 1.5 \cite{11} in comparison with standard SPR sensors.
Theoretical and experimental research of nanorod-mediated sensor \cite{13} with thin metal film and porous layer shows the sensitivity enhancement by twofold, but its sensitivity depends from the analyte refractive index value: with the refractive index increasing the prism replacing is needed.

Liquid crystal (LC) is an anisotropic medium so that the optical axis can be controlled by applied electric or magnetic fields \cite{14,15}. SPs with liquid crystals were investigated in different geometries \cite{16,17} and now LCs are a promising candidate for the development of active plasmonic devices \cite{171}. In this study, we consider a new sensor system with the additional nematic LC layer, which can be used for tuning sensor characteristics. Contrary to \cite{18}, an electric field is applied to the LC cell that causes the LC director reorientation. As a result, the LC director orientation and the LC dielectric tensor become inhomogeneous and dependent on the distance to the cell substrate.
In such a system, Fresnel equations based methods \cite{13,18} become inapplicable and matrix computational methods are preferable.
The study aims to theoretically investigate the effect of the LC reorientation on the spectral reflectance properties of the proposed sensor system.

\section{System scheme}
\begin{figure}[htb]
\centerline{\includegraphics[width=0.8\textwidth]{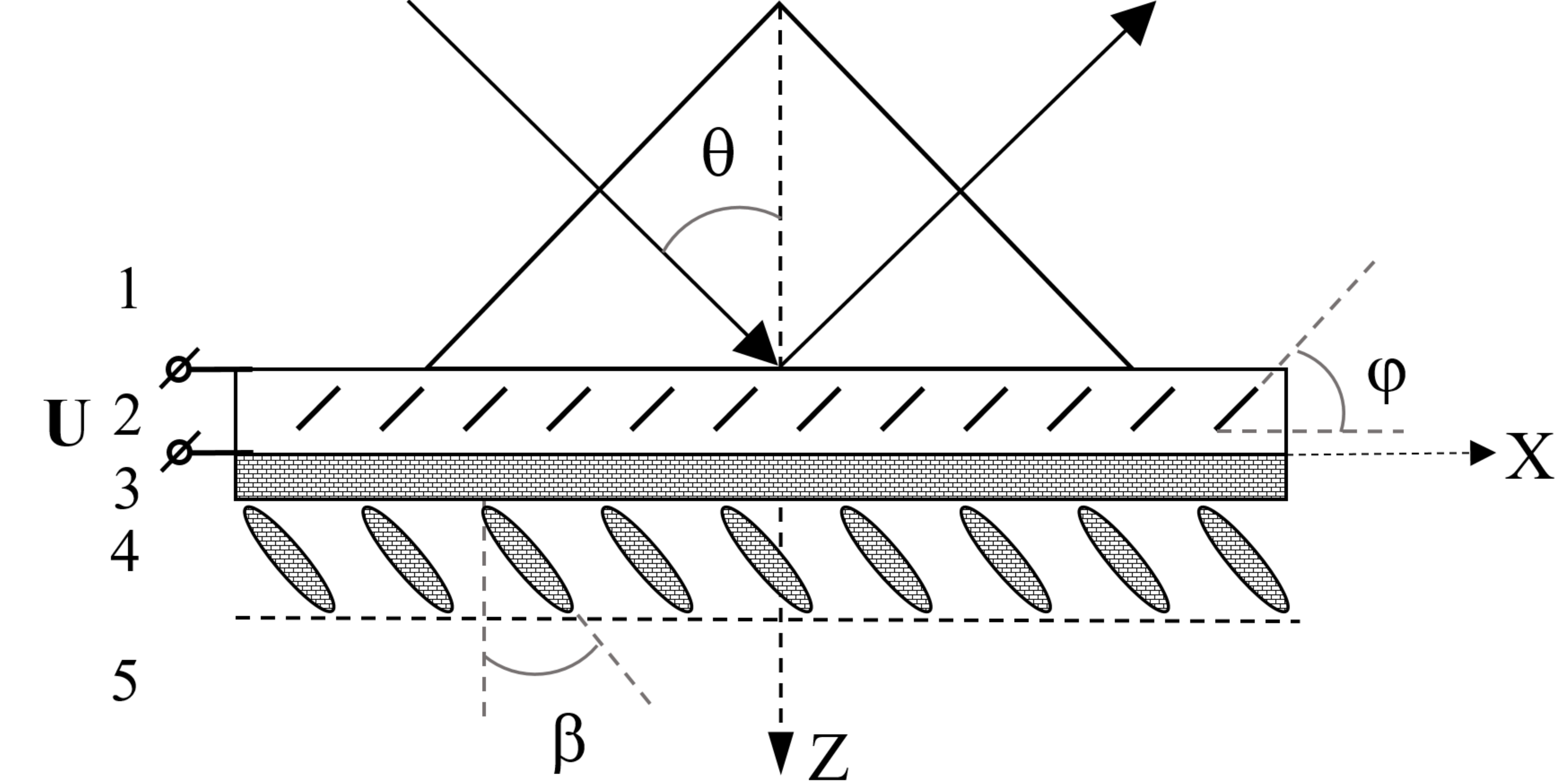}}
\caption{Geometry of the investigated system.}
\label{fig-1}
\end{figure}

The investigated system is based on Kretschman geometry that consists of a glass prism, a thin metal film and an analyte. Monochromatic p-polarized light is incident onto the interface between the glass prism and the metal film. The plasmon wave vector depends on the dielectric constants of the metal ($\varepsilon_\textrm{m}$) and analyte ($\varepsilon_{\textrm{s}}$) $k_\textrm{p}=\frac{2\piup}{\lambda}\sqrt{\frac{\varepsilon_\textrm{m}\varepsilon_{\textrm{s}}}{\varepsilon_\textrm{m}+\varepsilon_{\textrm{s}}}} $. When the tangential component of the incident light wave vector is equal to the plasmon wave vector, the plasmon is 
exited. The intensity of the reflected light is measured by a detector. The analyte refractive index can be calculated from the angular position of the resonance minimum. In \cite{13}, an anisotropic porous metal film was inserted between metal film and analyte for the sensitivity increase. We add a layer of nematic liquid crystal to change the sensor characteristics when it is already fabricated or during the measuring. The scheme of the investigated system is shown in figure~\ref{fig-1}~. It consists of a glass prism with refractive index $n_1$~(1), a nematic liquid crystal layer~(2), a thin metal (Ag) film~(3) and an anisotropic porous metallic film~(4) constituted by spheroidal nanoparticles, deposited so that their rotation axis is tilted at an angle $\beta$ to the $Z$-axis \cite{13} (for example, using the oblique-angle-deposition technique \cite{11}). An analyte forms the last layer~(5) and fills the spaces between nanoparticles which increases the sensitivity of the system \cite{13}. An external electric field is applied to the LC cell in the $Z$ direction. Monochromatic p-polarized light is incident at some angle $\theta$ onto the interface between the glass prism and the LC layer. The reflectance angular spectrum was investigated at different applied voltages.

\section{Theoretical model}

\subsection{Calculation method}

To calculate the propagation of the obliquely incident light in a system with the inhomogeneosly oriented LC layer (the LC director orientation depends on $Z$ coordinate), the Berreman $4\times4$ matrix method \cite{19, 20,201,202,203} was used. This method was developed for stratified anisotropic media \cite{203,204}, which is a more general approach compared to $2\times2$ matrix method and suitable for the oblique light incidence. The method is based on the solution of the Berreman equation that is the first order differential equation for the Berreman vector  $\boldsymbol{\psi}$
\begin{equation}
\frac{\rd \boldsymbol{\psi}(z)}{\rd z} =\ri k_0 \hat{Q}\boldsymbol{\psi}(z),
\label{eq:refname1}
\end{equation}
where $\boldsymbol{\psi}(z)=\left(E_x,\eta_0H_x,E_y,-\eta_0Hy\right)$, $\eta_0=\sqrt{\mu_0/\varepsilon_0}$, $k_0=\omega/c$ is the wave vector in vacuum, $\varepsilon_0$ are $\mu_0$ are vacuum dielectric and magnetic constants. The matrix $\hat{Q}$ depends on the components of the dielectric tensor $\hat{\varepsilon}$ and on the tangential component of the incident light wave vector. If the dielectric tensor does not change in the region from $z$ to $z+\Delta z$ then $\hat{Q}$ also does not change in this region and the solution to equation~(\ref{eq:refname1}) is of the form $\boldsymbol{\psi}\left(z+\Delta z\right)=\exp\left[\ri k_0\hat{Q}(z)\Delta z\right]\boldsymbol{\psi}(z)=\hat{P}\left(z,\Delta z\right)\boldsymbol{\psi}(z)$. The matrix exponent was calculated using faster Berreman method \cite{20,204} based on the  Cayley-Hamilton theorem. The LC cell was divided into 200 slabs with an approximately uniform director that, as it turned out, provides an acceptable accuracy in our calculations. The reflectance was calculated as a ratio between the intensities of the incident and the reflected waves using a technique from \cite{201}.

The following parameters were used for numerical simulation: the wavelength of the incident light $\lambda=632.8$ nm, the glass refractive index 1.51 and 1.57, the metal film thickness $d_\textrm{f}=40$ nm, the diameter of nanoparticles $D=30$ nm and length $l=10$ nm (oblate spheroid), the angle between rotation axis and $Z$-axis $\beta=73^\circ$, the volume fraction $f_{\text{m}}=0.4$.
\subsection{Dielectric tensors }
To calculate the reflectance using the Berreman method one needs to know the dielectric function of each layer. For silver dielectric constant, we used the approximation $\varepsilon (\omega)=1-\frac{\omega^2_\textrm{p}}{\omega(\omega+\ri\gamma)}+\frac{f\omega^2_\textrm{b}}{\omega^2_\textrm{b}-\omega^2+\ri\omega\Gamma_\textrm{b}}$,
where the last term is an approximation by Lorentzian tail of the contribution from interband electron transitions \cite{21}, $\omega_\textrm{p}$ is the plasma frequency, $\gamma$ is the relaxation constant. This approximation is more general than the one used in \cite{13}. For silver film, it gives a result close to the Drude formula $\varepsilon_{\text{m}}=-18.37+\ri 0.47$. For nanoparticles, the surface effect on the electron mean free path ($L$) is rather strong and $\gamma (L)$ is of the form $\gamma (L)=\gamma_0+A{v_\textrm{f}}/{L}$, where $\gamma_0$ is the volume decay constant, $v_\textrm{f}$ is the Fermi velocity of electrons, $A$ is a parameter of the order of  unity that accounts for the details of the scattering process. The electron mean free path $L$ was calculated as in \cite{22}.

The anisotropic porous metal film consists of the array of spheroidal nanoparticles. Since the rotation axis is tilted at the angle $\beta$ to the $Z$-axis, the dielectric tensor of the layer is of the form
\begin{equation}
\hat{\varepsilon}_4=\begin{pmatrix} \varepsilon_{x'} \cos^2\beta+ \varepsilon_{z'} \sin^2\beta& 0  &(\varepsilon_{z'}-\varepsilon_{x'}) \sin\beta \cos\beta\\ 0 & \varepsilon_{y'}  & 0\\ (\varepsilon_{z'}-\varepsilon_{x'})\sin\beta\cos\beta & 0 & \varepsilon_{z'} \cos^2\beta+ \varepsilon_{x'} \sin^2\beta \end{pmatrix},
\label{eq:refname3}
\end{equation}
where $\varepsilon_{x'},\varepsilon_{y'},\varepsilon_{z'} $ are the principal values. The analyte fills the spaces between nanoparticles and the formed layer is inhomogeneous. Such a structure can be considered as a homogeneous optical medium with an effective dielectric constant that is different from dielectric constants of constituting materials. In the Maxwell Garnett model, the effective dielectric constant $\varepsilon$ of the system with a homogeneous host and spherical inclusions can be obtained from the equation  $\frac{\varepsilon-\varepsilon_2}{\varepsilon+2\varepsilon_2}=f_1\frac{\varepsilon_1-\varepsilon_2}{\varepsilon_1+2\varepsilon_2}$, where $\varepsilon_2, \varepsilon_1$ are host and inclusions dielectric constants, respectively, $f_1$ is the inclusions volume fraction \cite{23}. The mixing formula is modified for the case of the ellipsoidal particles. Considering the metal as the host \cite{13}, we used the modified formula that is of the form
\begin{equation}
\frac{\varepsilon_i-\varepsilon_\textrm{m}}{\varepsilon_\textrm{m}+L_i(\varepsilon_i-\varepsilon_\textrm{m})}=
(1-f_\textrm{m})\frac{\varepsilon_{\textrm{s}}-\varepsilon_\textrm{m}}{\varepsilon_\textrm{m}+L_i(\varepsilon_{\textrm{s}}-\varepsilon_\textrm{m})}\,,
\label{eq:refname4}
\end{equation}
where $i=x',y',z'$, $\varepsilon_\textrm{m}$ and $\varepsilon_{\textrm{s}}$ are metal and analyte dielectric constants, $f_\textrm{m}$ is the metal volume fraction, $L_i$ is the ellipsoid depolarization factor \cite{24}.

\subsection{Equations for the LC director  }
The liquid crystal layer is located between the glass prism and the metal film in the system under consideration. The LC director orientation at top and bottom substrates are parallel to the $X$-axis. When the applied voltage is higher than the electric Frederiks transition threshold, the director rotates in the $XZ$-plane. Its orientation can be described by the unit vector $\mathbf{n}=\cos\varphi(z)\mathbf{e}_x+\sin\varphi(z)\mathbf{e}_z$, where $\varphi(z)$ is the rotation angle. Components of the LC dielectric tensor depend on the director orientation
\begin{equation}
\hat{\varepsilon}_{\textrm{LC}}=\begin{pmatrix} n_\textrm{e}^2 \cos^2\varphi+ n_\textrm{o}^2 \sin^2\varphi& 0  &\big(n_\textrm{e}^2-n_\textrm{o}^2\big) \sin\varphi \cos\varphi\\ 0 & n_0^2  & 0\\ \big(n_\textrm{e}^2-n_\textrm{o}^2\big) \sin\varphi \cos\varphi & 0 & n_\textrm{e}^2 \sin^2\varphi+ n_\textrm{o}^2 \cos^2\varphi \end{pmatrix},
\label{eq:refname5}
\end{equation}
where $n_\textrm{o}$ and $n_\textrm{e}$ are ordinary and extraordinary refractive indices. The director orientation under the applied voltage can be found by minimizing the LC free energy functional. The LC free energy density has two components: elastic $f_{\textrm{elastic}}=\frac{1}{2}K_{11}\cos^2\varphi(\varphi')^2+\frac{1}{2}K_{33}\sin^2\varphi(\varphi')^2 $ and electric $f_{\textrm{electric}}=-\frac{1}{2}\varepsilon_0\Delta\varepsilon(\mathbf{E}\cdot\mathbf{n})^2=-\frac{1}{2}\varepsilon_0\Delta\varepsilon E^2\sin^2\varphi$, where
$K_{11}$ and $K_{33}$ are Frank elastic constants, $\Delta\varepsilon$ is a static dielectric anisotropy, $E$ is the electric field in the LC layer. By varying the free energy functional, the second order differential equation was obtained
\begin{equation}
\big(K_{11}\cos^2\varphi+K_{33}\sin^2\varphi\big)\varphi''+\big(K_{33}-K_{11}\big)\sin\varphi\cos\varphi(\varphi')^2+\varepsilon_0\Delta\varepsilon E^2\sin\varphi\cos\varphi=0.
\label{eq:refname6}
\end{equation}
This equation must be accompanied by boundary conditions. We considered two variants of boundary conditions: 1) strong anchoring at both surfaces (symmetric cell) $\varphi(z=0)=\varphi(z=h)=0$ with the threshold voltage $U_{\text{th}}=\piup\sqrt{\frac{K_{11}}{\varepsilon_0\Delta\varepsilon}}$\,; 2) strong anchoring at the top surface and very weak anchoring at the bottom surface (non-symmetric cell) $\varphi(z=0)$, $\varphi'(z=h)=0$ with the threshold voltage $U_{\text{th}}NS=\frac{\piup}{2}\sqrt{\frac{K_{11}}{\varepsilon_0\Delta\varepsilon}}$\,. The electric field from equation~(\ref{eq:refname6}) obeys the equation $\left(\pmb{\nabla}\cdot\mathbf{D}\right)=0$ that yields
\begin{eqnarray}
\frac{\rd}{\rd z}(\varepsilon_0\varepsilon_{zz}E)=\varepsilon_0\frac{\rd}{\rd z} \Big\{\bigl[\varepsilon_{\bot}+\Delta\varepsilon \sin^2 \varphi(z)\bigr] E \Big\}=0\,, 
\label{eq:refname7}
\end{eqnarray}
where $\varepsilon_{\bot}$ is the perpendicular component of the static dielectric tensor. It is convenient to introduce the electric field potential $\mathbf{E}=-\pmb{\nabla}\Phi$. Then the potential satisfies the following equation and boundary conditions
\begin{eqnarray}
\frac{\rd}{\rd z} \bigg\{\bigl[\varepsilon_{\bot}+\Delta\varepsilon\sin^2\varphi(z) \bigr] \frac{\rd \Phi}{\rd z} \bigg\}=0\,,
\label{eq:refname8}
\end{eqnarray}
\[\Phi(0)=0, \qquad \Phi(h)=U.\]
Table~\ref{tbl-LCp} shows the parameters of the chosen nematic LCs 5CB and E44. The LC cell thickness is $h=5$~\textmu{}m.

\begin{table}[htb]
\caption{LC parameters \cite{15,25,26} and threshold voltages.}
\label{tbl-LCp}
\vspace{1ex}
\begin{center}
\renewcommand{\arraystretch}{0}
\begin{tabular}{|c|c|c|c|c|c|c|c|c|}
\hline\hline
LC&$K_{11}$,N&$K_{33}$,N&$\Delta\varepsilon$&$\varepsilon_{\bot}$&$n_{\textrm{o}}$&$n_{\textrm{e}}$&$U_{\text{th}}$,V&$U_{\text{th}}NS$,V\strut\\
\hline\hline
\rule{0pt}{2pt}&&&&&&&&\\
5CB&$0.64\cdot10^{-11}$&$0.10\cdot10^{-10}$&13&6.7&1.5319&1.7060&0.7408&0.3704\strut\\
\hline
\rule{0pt}{2pt}&&&&&&&&\\
E44&$0.155\cdot10^{-10}$&$0.28\cdot10^{-10}$&16.8&5.2&1.5239&1.7753&1.0141&0.50705\strut\\
\hline\hline
\end{tabular}
\renewcommand{\arraystretch}{1}
\end{center}
\end{table}

\section{Result and discussions }
\subsection{The prism refractive index 1.51}

\begin{figure}[!b]
\includegraphics[width=0.5\textwidth]{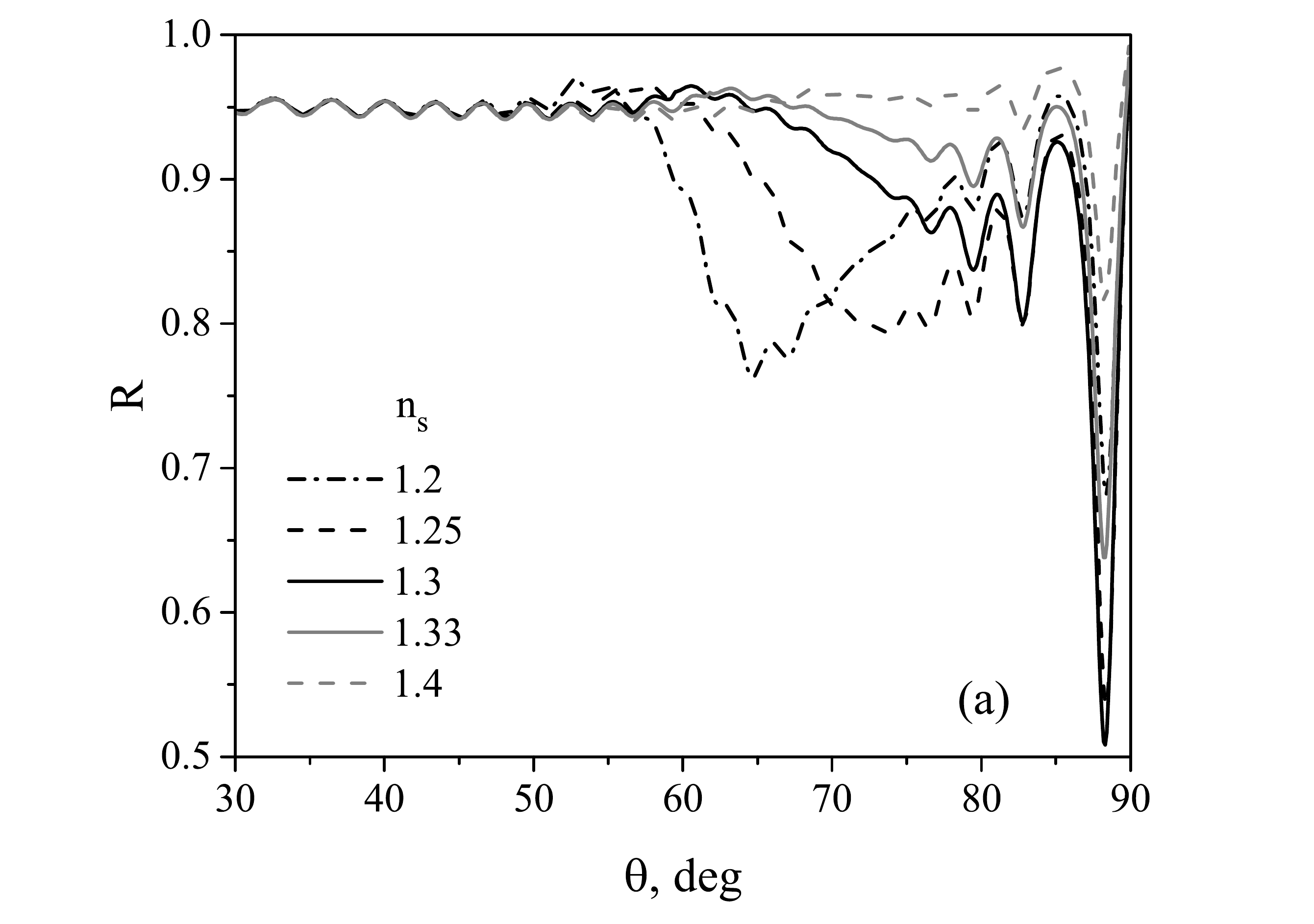}\includegraphics[width=0.5\textwidth]{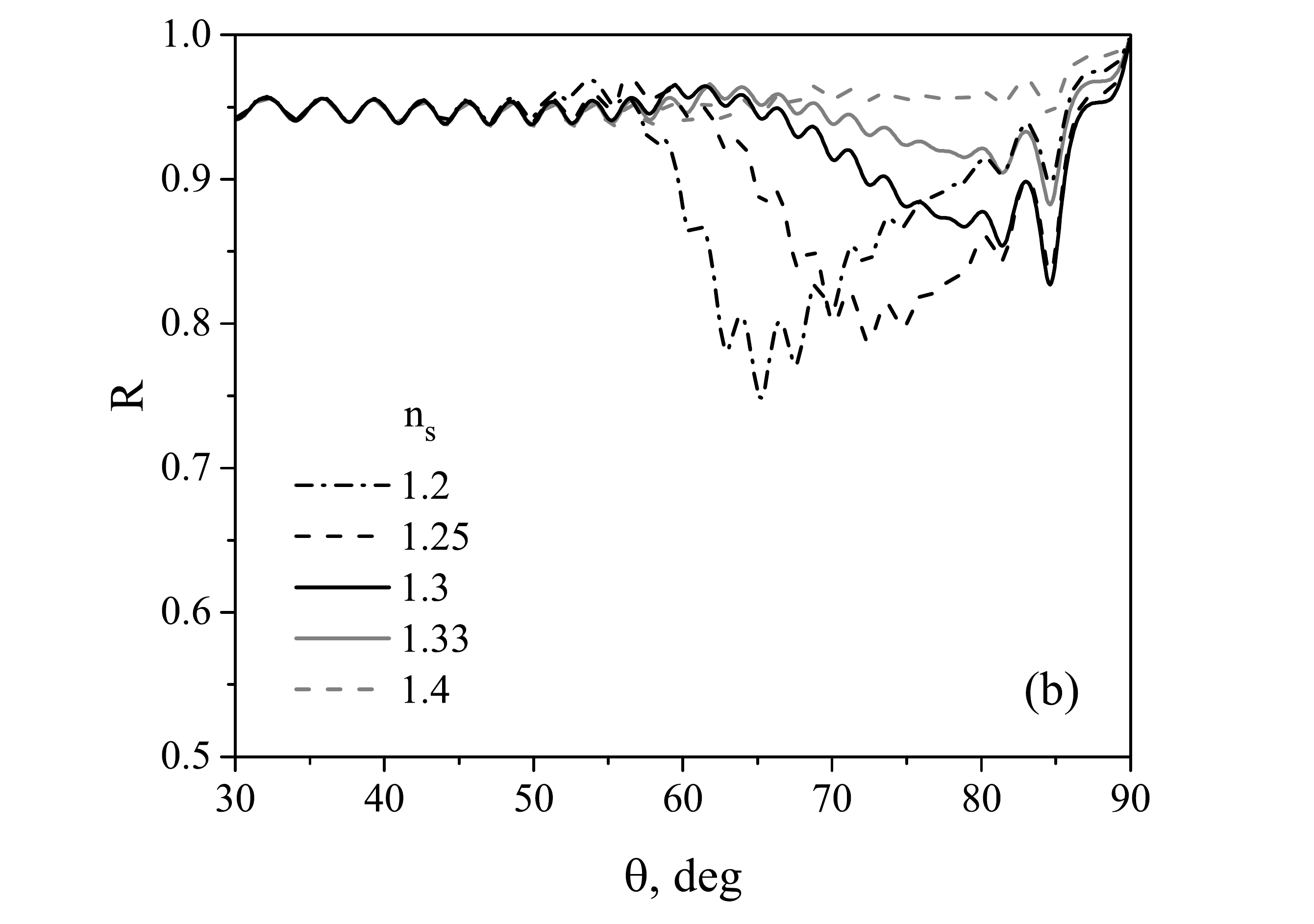}
\caption{Reflectance $R$ versus the light incident angle $\theta$ without the LC director reorientation (planar geometry) for 5CB (a) and E44 (b) at different values of the analyte refractive index $n_{\text{s}}$, $n_1=1.51$.}
\label{fig-2}
\end{figure}

\begin{figure}[!b]
	\centerline{\includegraphics[width=0.5\textwidth]{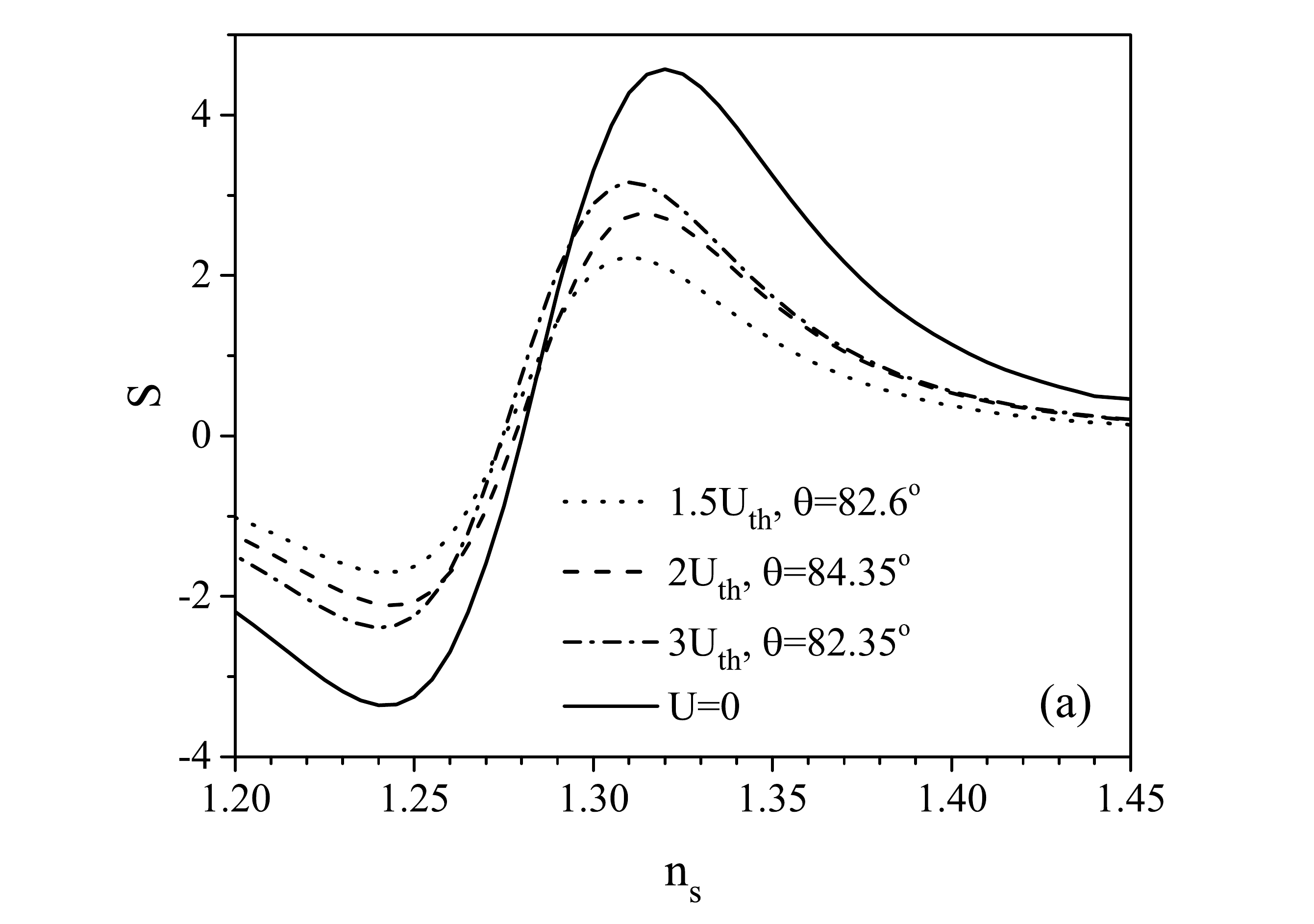}\includegraphics[width=0.5\textwidth]{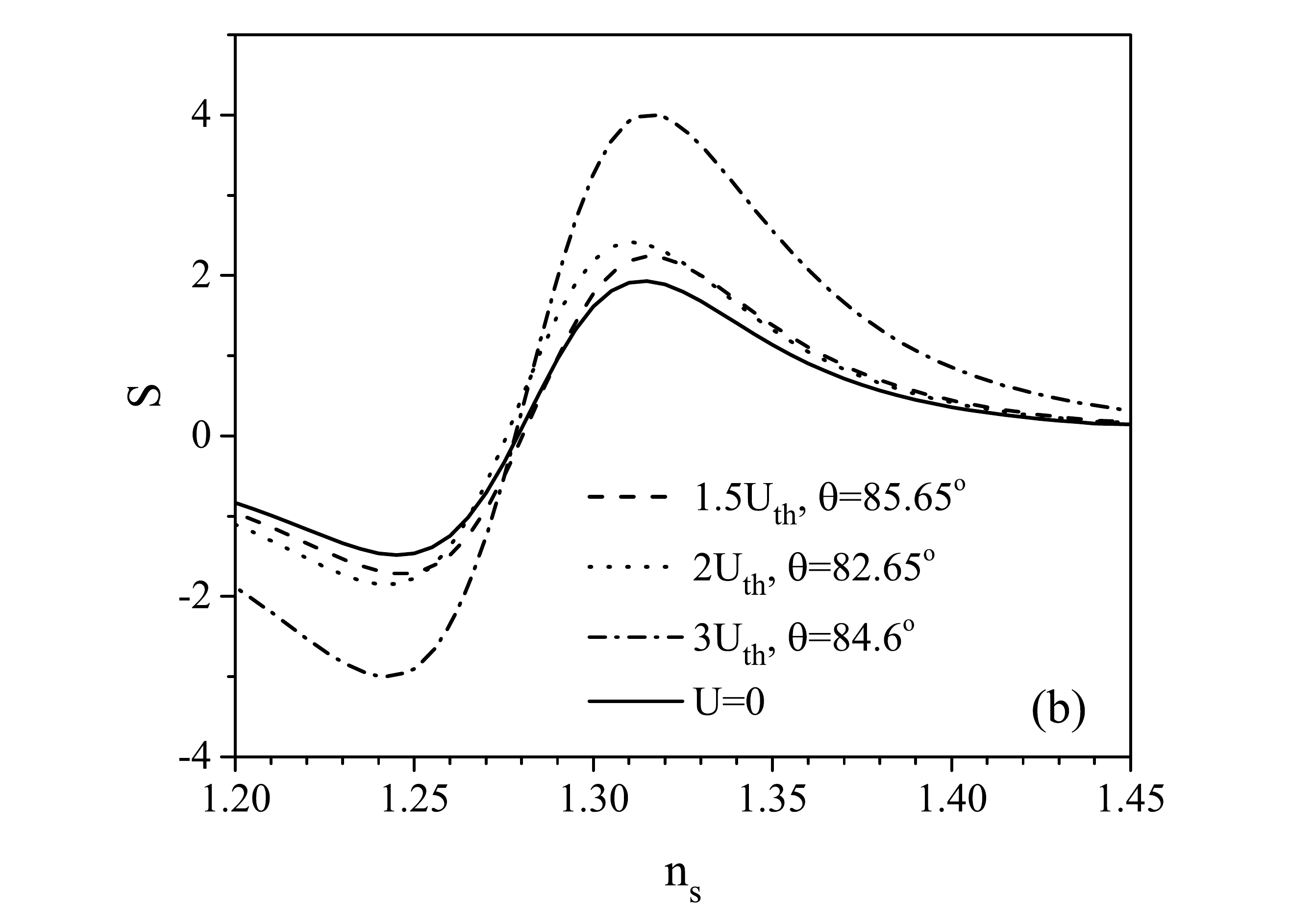}}
	\caption{Sensitivity $S$ versus the analyte refractive index $n_{\text{s}}$ for 5CB (a) and E44 (b) without reorientation ($U=0$) and at voltages 1.5, 2 and 3$U_{\text{th}}$, $n_1=1.51$.}
	\label{fig-5}
\end{figure}

Figure~\ref{fig-2} shows the light reflectance $R$ without the LC director reorientation (planar geometry) for 5CB (a) and E44 (b) at different values of the analyte refractive index $n_{\text{s}}$; the prism refractive index is $n_1=1.51$.
 Similar to the case \cite{18} there are dips in the curves $R(\theta)$, that are caused by mixing of the surface plasmons mode with the half-leak modes propagating in the LC layer. Plasmon resonance similar to the one in \cite{13} is observed at small $n_{\text{s}}$ values. There is also a dip on the right side of figure~\ref{fig-2}, its position is fixed for the particular LC and the depth depends on the analyte refractive index $n_{\text{s}}$. At the incident angle that corresponds to the dip minimum, the analyte refractive index can be found by using the reflectance value. The sensor sensitivity $S$ to the analyte refractive index change is defined as the ratio between the reflectance change and the analyte refractive index change $\rd R_{\text{min}}/\rd n_{\text{s}}$. For example, in figure~\ref{fig-2}~(a) (5CB) for $\theta_{\text{min}}=88.3^\circ$ the calculated sensitivity for $n_{\text{s}}=1.33-1.34$ is $S=4.11$; in figure~\ref{fig-2}~(b) (E44) for $\theta_{\text{min}}=84.6^\circ$ $S=1.55$. Sensitivity dependence on $n_{\text{s}}$ is shown in figure~\ref{fig-5}. It has a maximum at some $n_{\text{s}}$ value, but when $n_{\text{s}}$ increases, sensitivity decreases. The maximum sensitivity for 5CB $S_{\text{max}}=4.57$ is achieved at $n_{\text{s}}=1.32$, for E44 $S_{\text{max}}=1.93$ at $n_{\text{s}}=1.315$.

\begin{figure}[!t]
	\centerline{\includegraphics[width=0.51\textwidth]{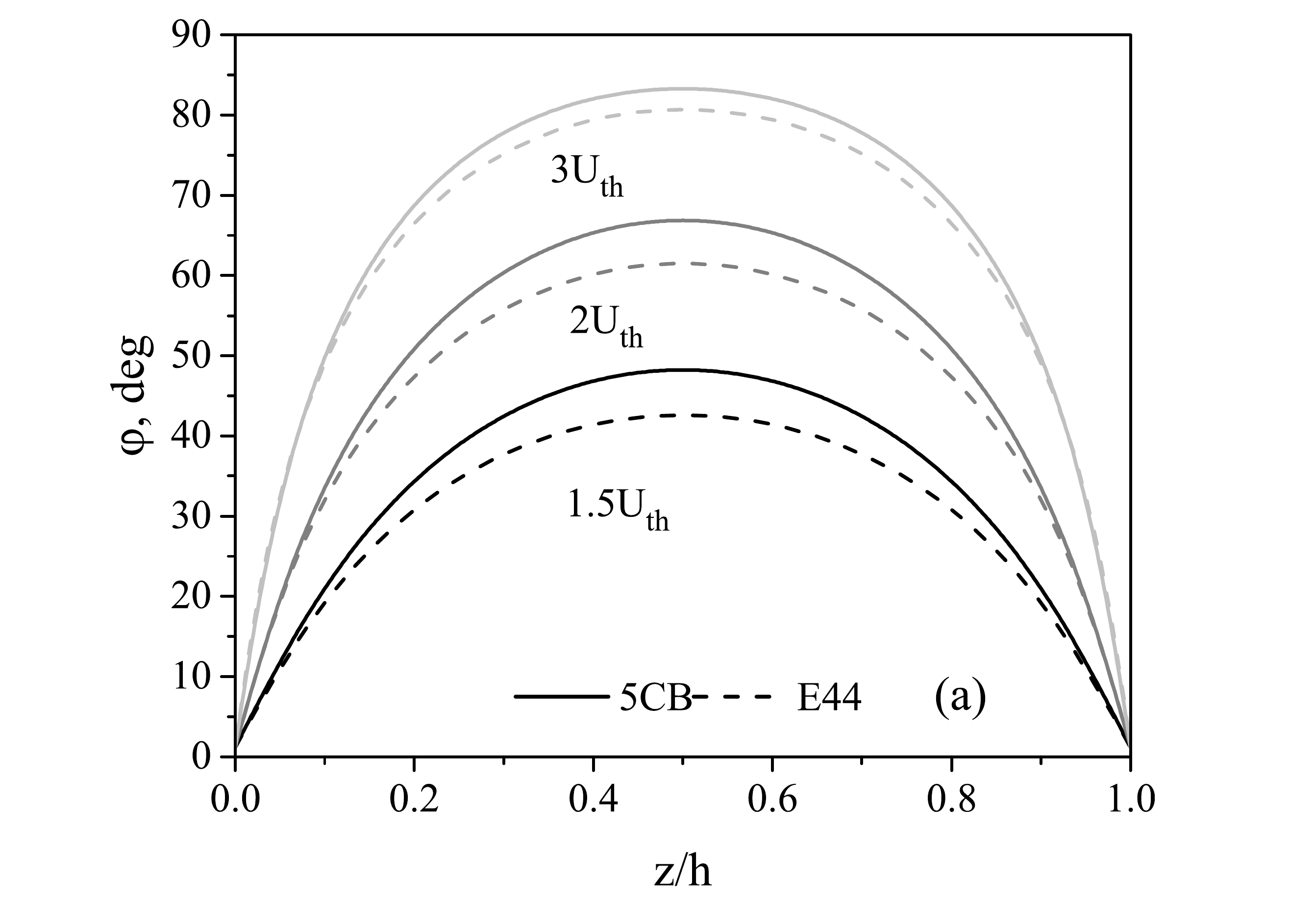}\includegraphics[width=0.51\textwidth]{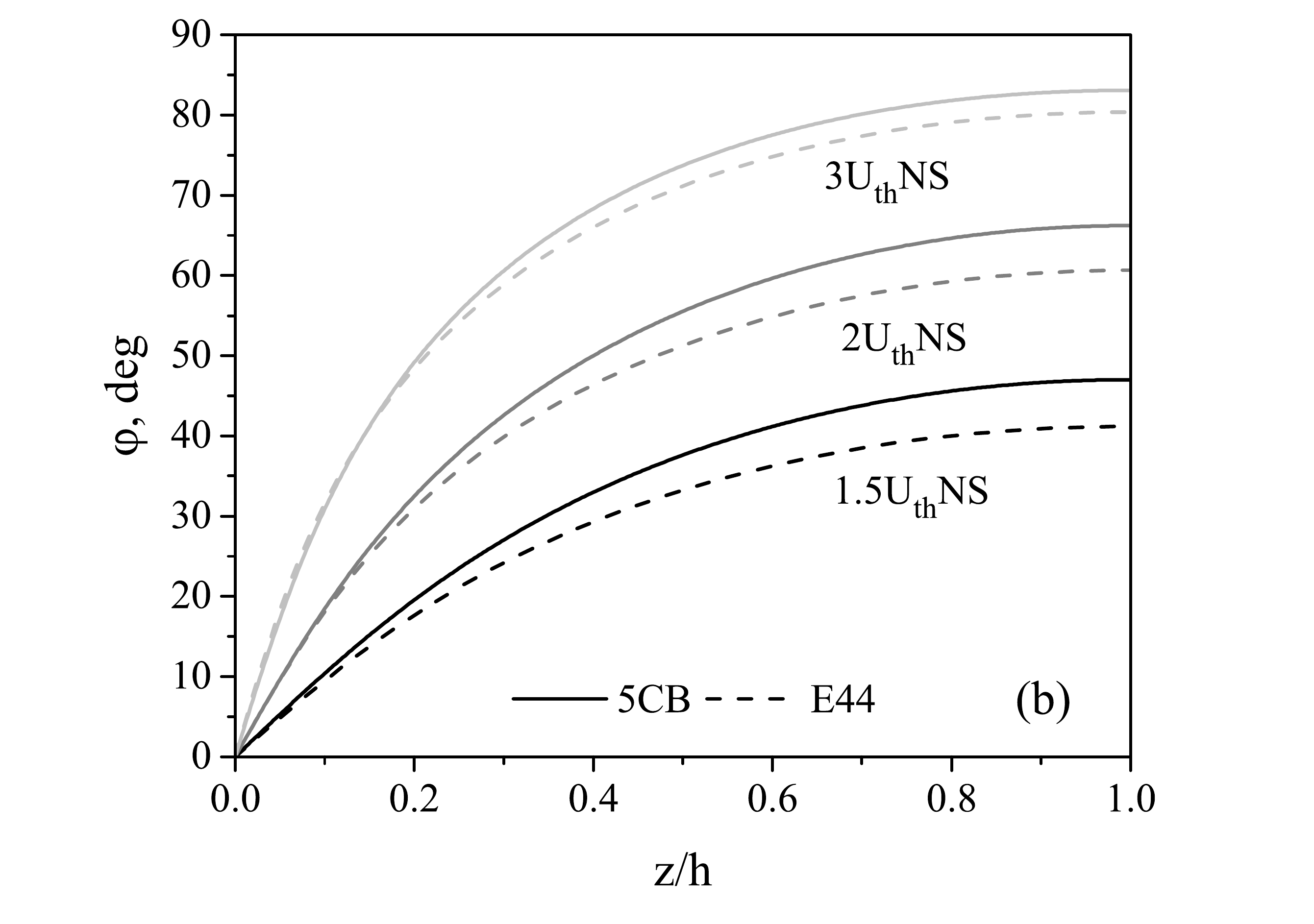}}
	\caption{Director profiles under the applied voltage: strong anchoring at both surfaces (a), strong anchoring at the top surface and very weak anchoring at the bottom surface (b).}
	\label{fig-3}
\end{figure}

\begin{figure}[!t]
	\centerline{\includegraphics[width=0.51\textwidth]{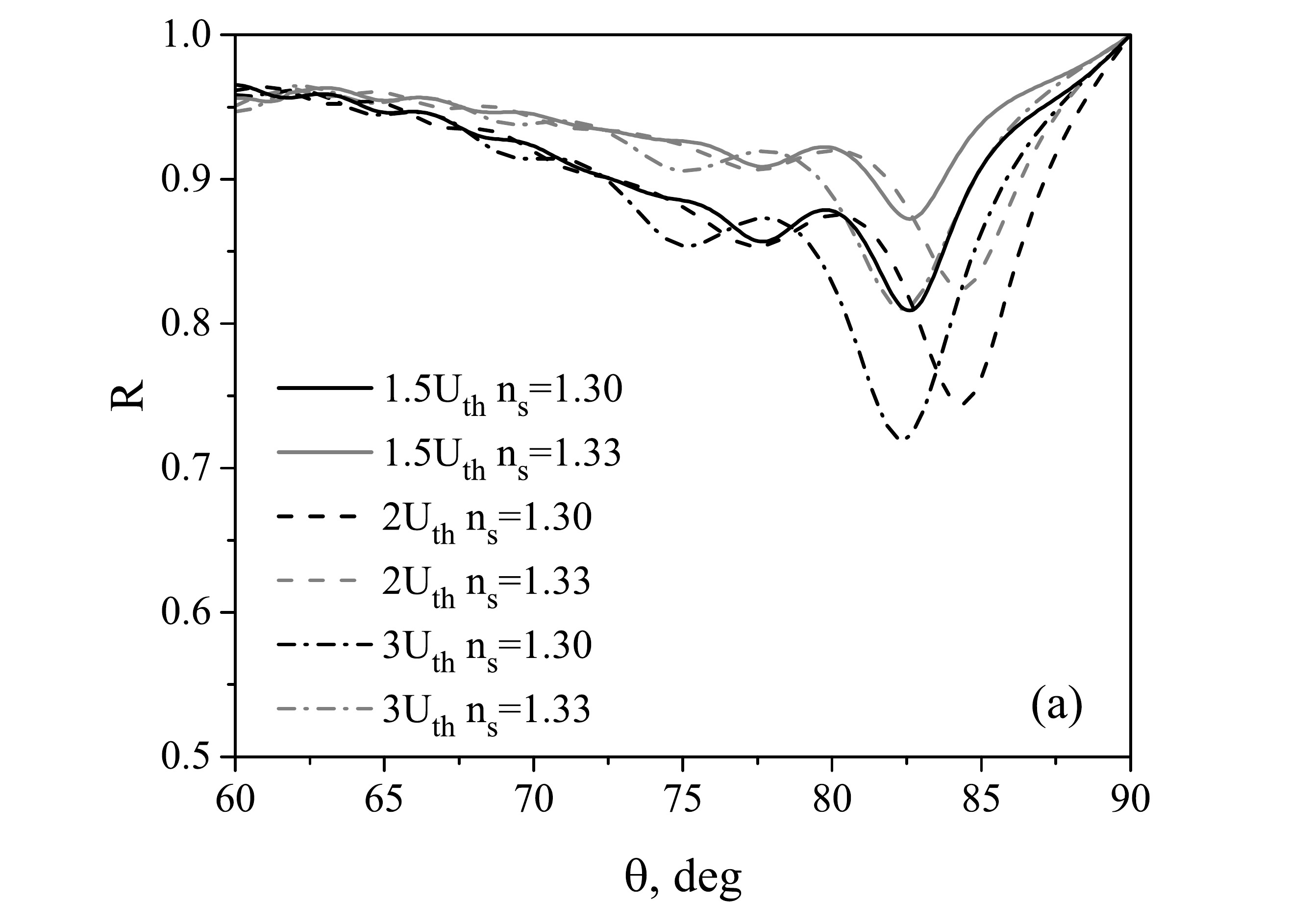}\includegraphics[width=0.51\textwidth]{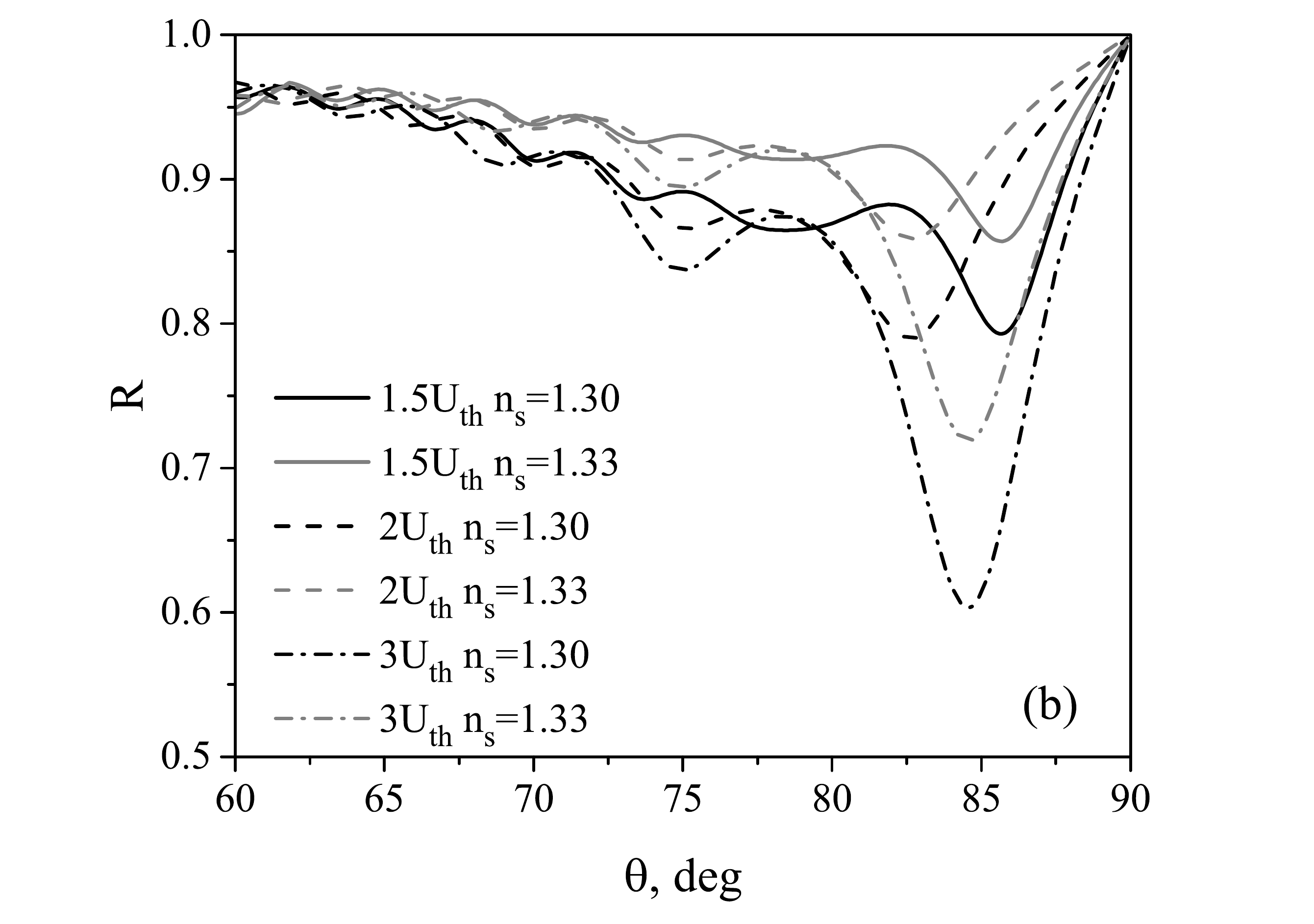}}
	\caption{Reflectance $R$ versus the light incident angle $\theta$ for 5CB (a) and E44 (b) at voltages 1.5, 2 and 3$U_{\text{th}}$, $n_1=1.51$, $n_{\text{s}}=1.3$, 1.33.}
	\label{fig-4}
\end{figure}

In order to investigate the effect of the LC reorientation, director profiles were calculated for different values of the applied voltage. Profiles obtained for two cases of boundary conditions are shown in figure~\ref{fig-3}. Figure~\ref{fig-4} shows the reflectance for symmetric LC cell at voltages 1.5, 2 and $3U_{\text{th}}$.
 These curves are similar to the curves in figure~\ref{fig-2}, but positions and depths of the last dips change depending on the applied voltage. For 5CB [figure~\ref{fig-4}~(a)], the dips shift to smaller incident angles and their depths decrease. In the case of E44 [figure~\ref{fig-4}~(b)], dips become deeper and stay in the same range of the incident angles. Sensitivity dependences on the analyte refractive index at voltages 1.5, 2 and 3$U_{\text{th}}$ are shown in figure~\ref{fig-5}~. For 5CB at voltage 1.5$U_{\text{th}}$, sensitivity has a maximum $S_{\text{max}}=2.23$ at $n_{\text{s}}=1.31$, for the other voltages, the maximum sensitivities are as follows: $S_{\text{max}}=2.78$ at $n_{\text{s}}=1.315$ ($U=2U_{\text{th}}$) and $S_{\text{max}}=3.16$ at $n_{\text{s}}=1.31$ ($U=3U_{\text{th}}$). For E44, the maximum sensitivities are as follows: $S_{\text{max}}=2.23$ at $n_{\text{s}}=1.315$ ($U=1.5U_{\text{th}}$), $S_{\text{max}}=2.41$ at $n_{\text{s}}=1.31$ ($U=2U_{\text{th}}$), $S_{\text{max}}=4.02$ at $n_{\text{s}}=1.315$ ($U=3U_{\text{th}}$). For E44, reorientation causes the sensitivity increase, but for both LC, reorientation does not have a significant impact on the $n_{\text{s}}$ value at which the sensitivity has a maximum (figure~\ref{fig-5}). In the case of the second boundary conditions, reflectance and sensitivity curves look similar but the maximum sensitivities are smaller.
\subsection{The prism refractive index 1.57}
\begin{figure}[!t]
\centerline{\includegraphics[width=0.51\textwidth]{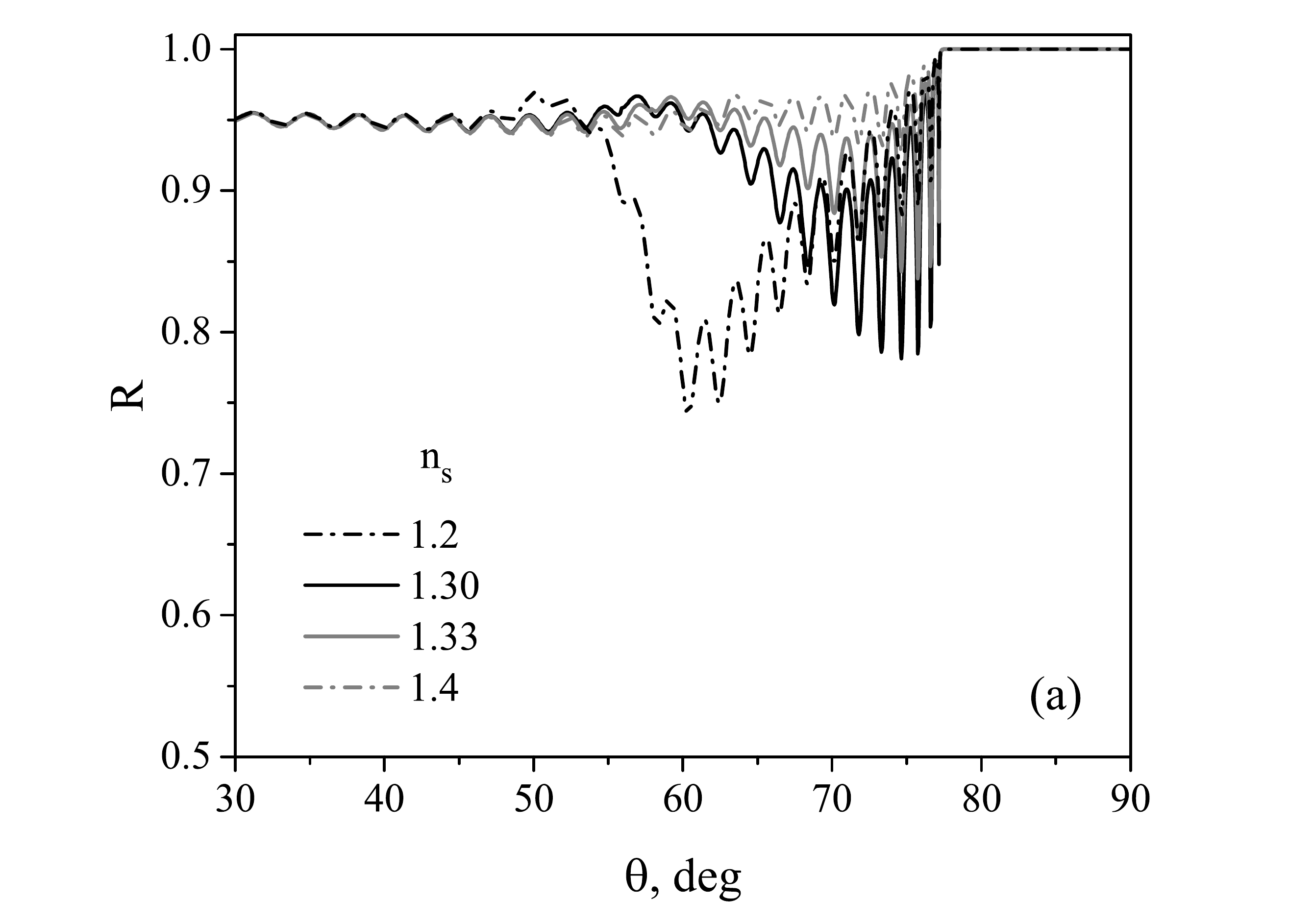}\includegraphics[width=0.51\textwidth]{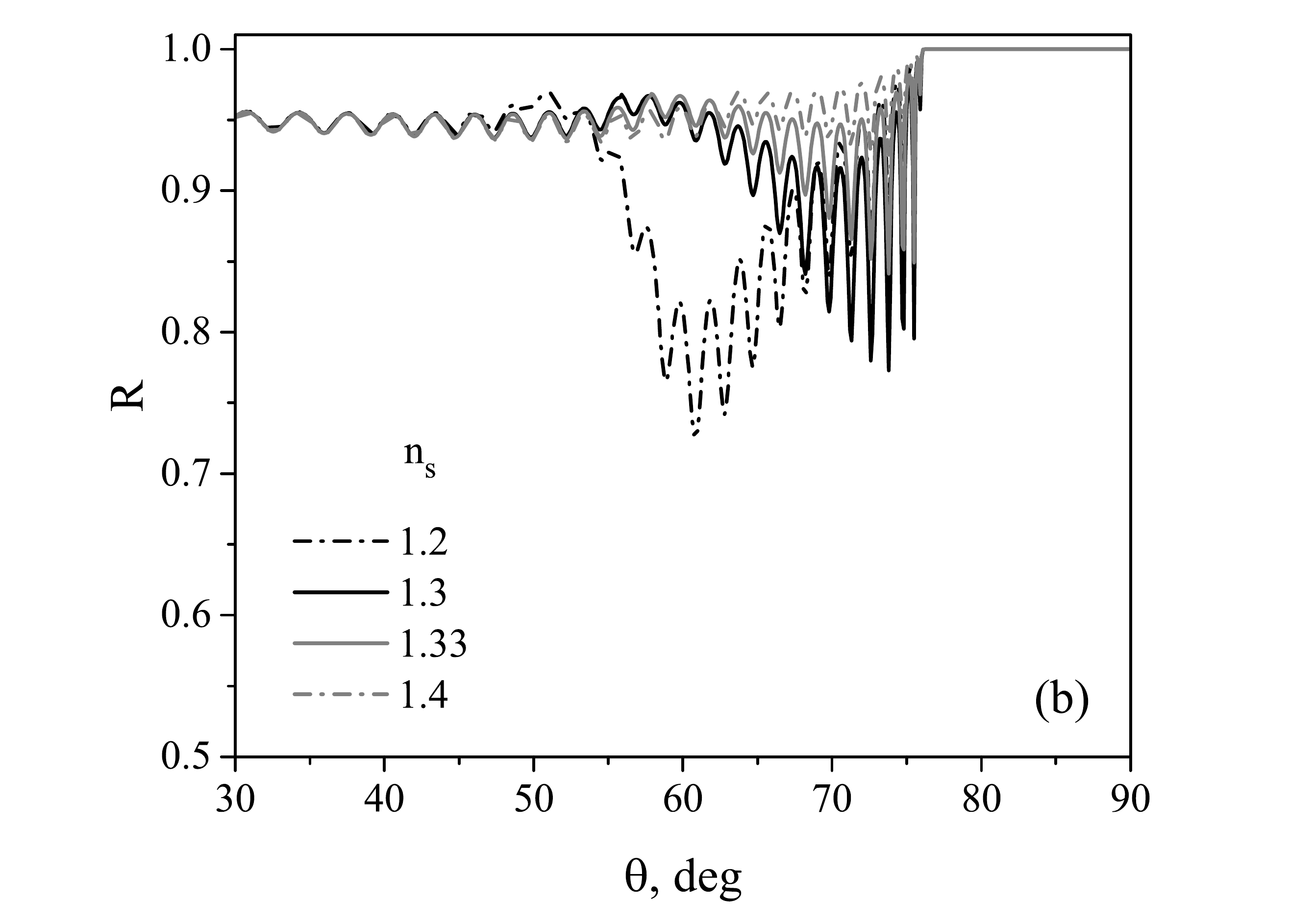}}
\caption{Reflectance $R$ versus the light incident angle $\theta$ for planar orientation for 5CB (a) and E44 (b) and different values of the analyte refractive index $n_{\text{s}}$, $n_1=1.57$.}
\label{fig-6}
\end{figure}

The prism refractive index 1.51 is smaller than ordinary ($n_{\text{o}}$) and extraordinary ($n_{\text{e}}$) refractive indices of the considered LCs. Since in this case reorientation does not have a significant impact on the maximum sensitivity position, the reflectance was investigated at the prism refractive index value between $n_{\text{o}}$ and $n_{\text{e}}$. Under this condition, LC reorientation changes the ratio between the prism and the LC layer refractive indices. At some LC orientation, total internal reflection can occur which is analogy to the Otto geometry \cite{3,15}. The reflectance for planar LC orientation at $n_1=1.57$ is shown in figure~\ref{fig-6}.
After some critical angle ($\theta_{\text{cr}}=77.3^\circ$ for 5CB and $\theta_{\text{cr}}=76.1^\circ $ for E44), the reflectance reaches a maximum and becomes constant $R=1$. When $n_1$ increases, the critical angle decreases. Since the reflectance dips are narrow and have a small depth, we consider this case unsuitable for $n_{\text{s}}$ measuring.

\begin{figure}[!t]
\centerline{\includegraphics[width=0.51\textwidth]{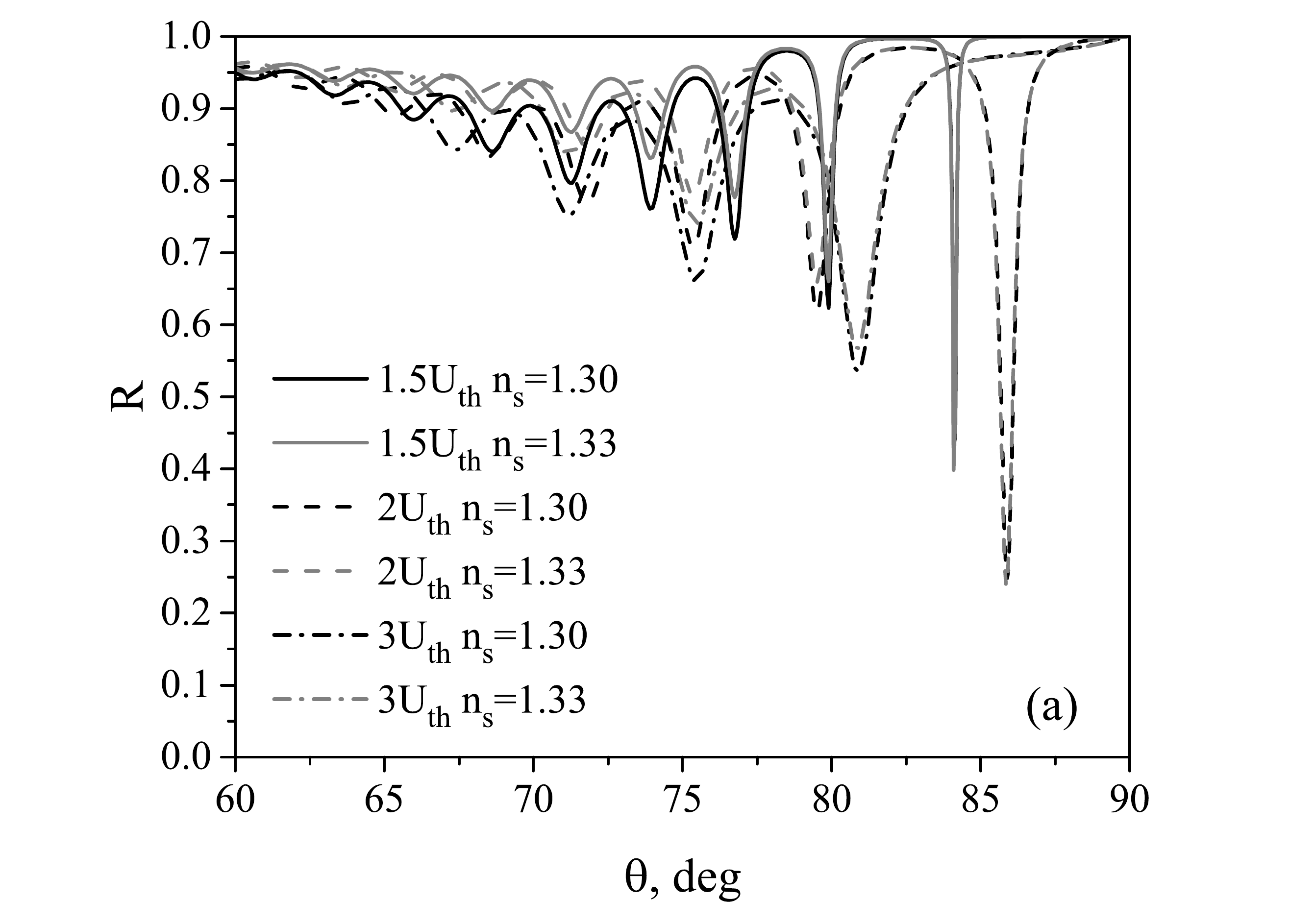}\includegraphics[width=0.51\textwidth]{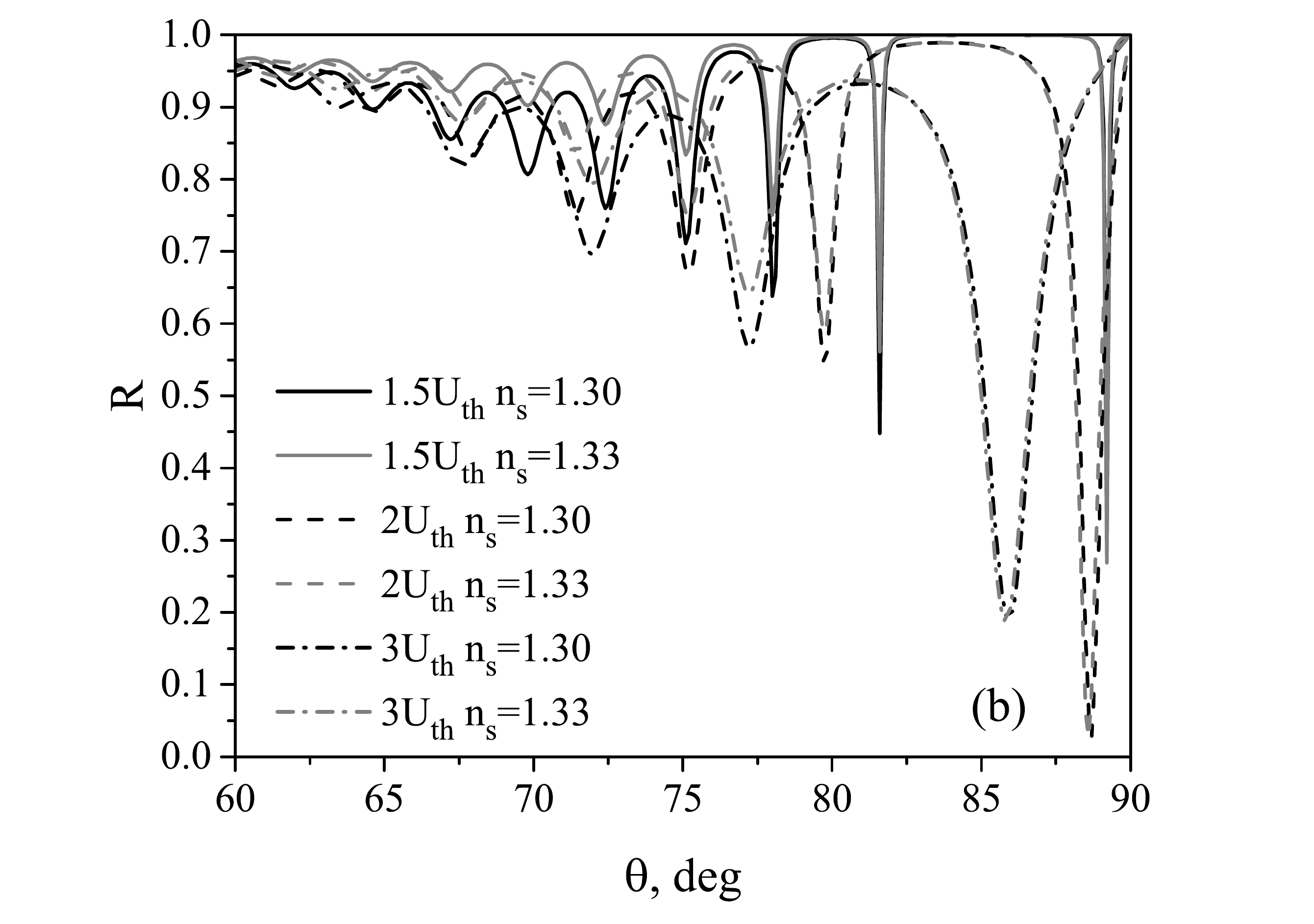}}
\caption{Reflectance $R$ versus the light incident angle $\theta$ for 5CB (a) and E44 (b) at voltages 1.5, 2 and 3$U_{\text{th}}$, $n_1=1.57$, $n_{\text{s}}=1.3,1.33$.}
\label{fig-7}
\end{figure}

\begin{figure}[!t]
\centerline{\includegraphics[width=0.5\textwidth]{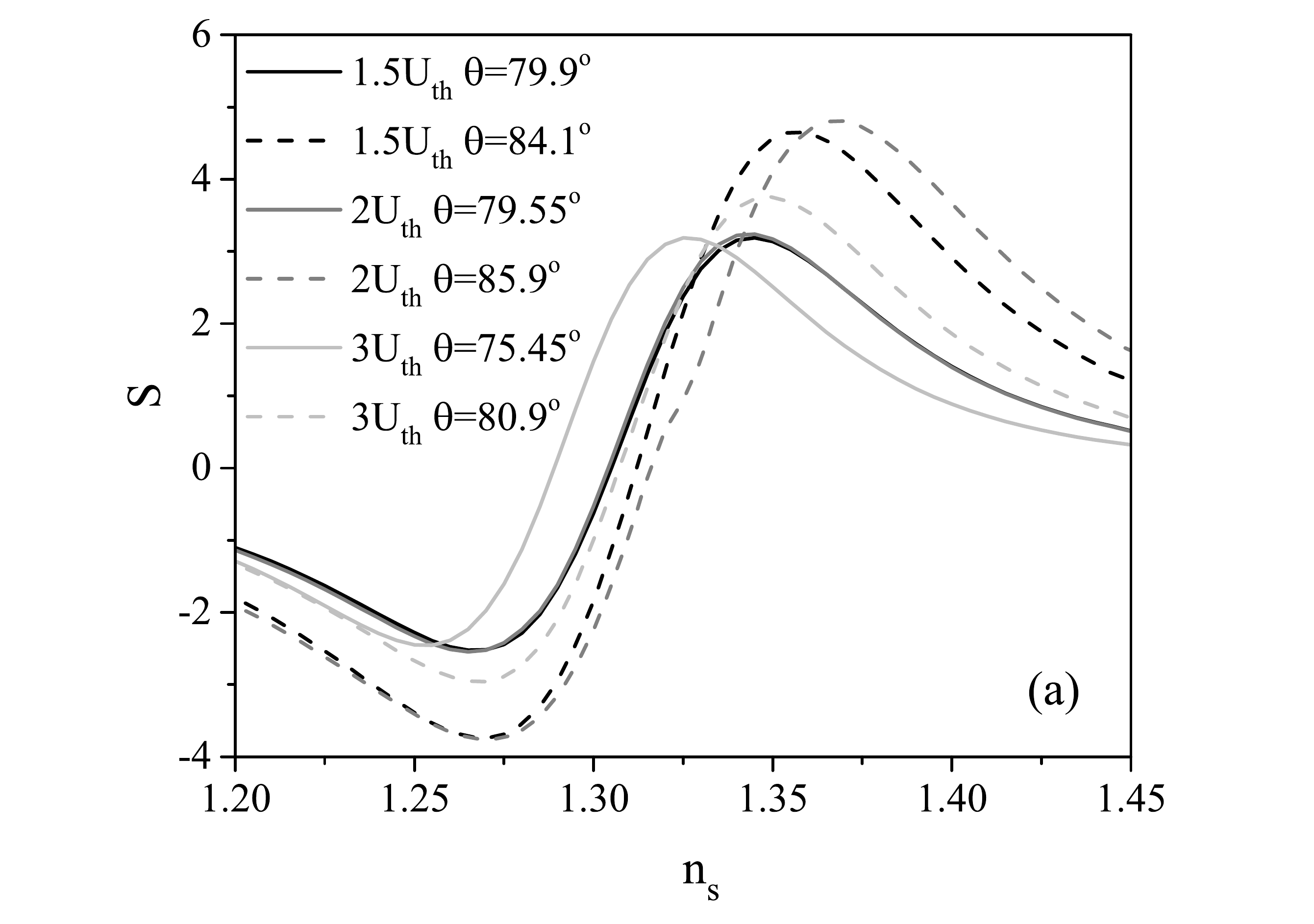}\includegraphics[width=0.5\textwidth]{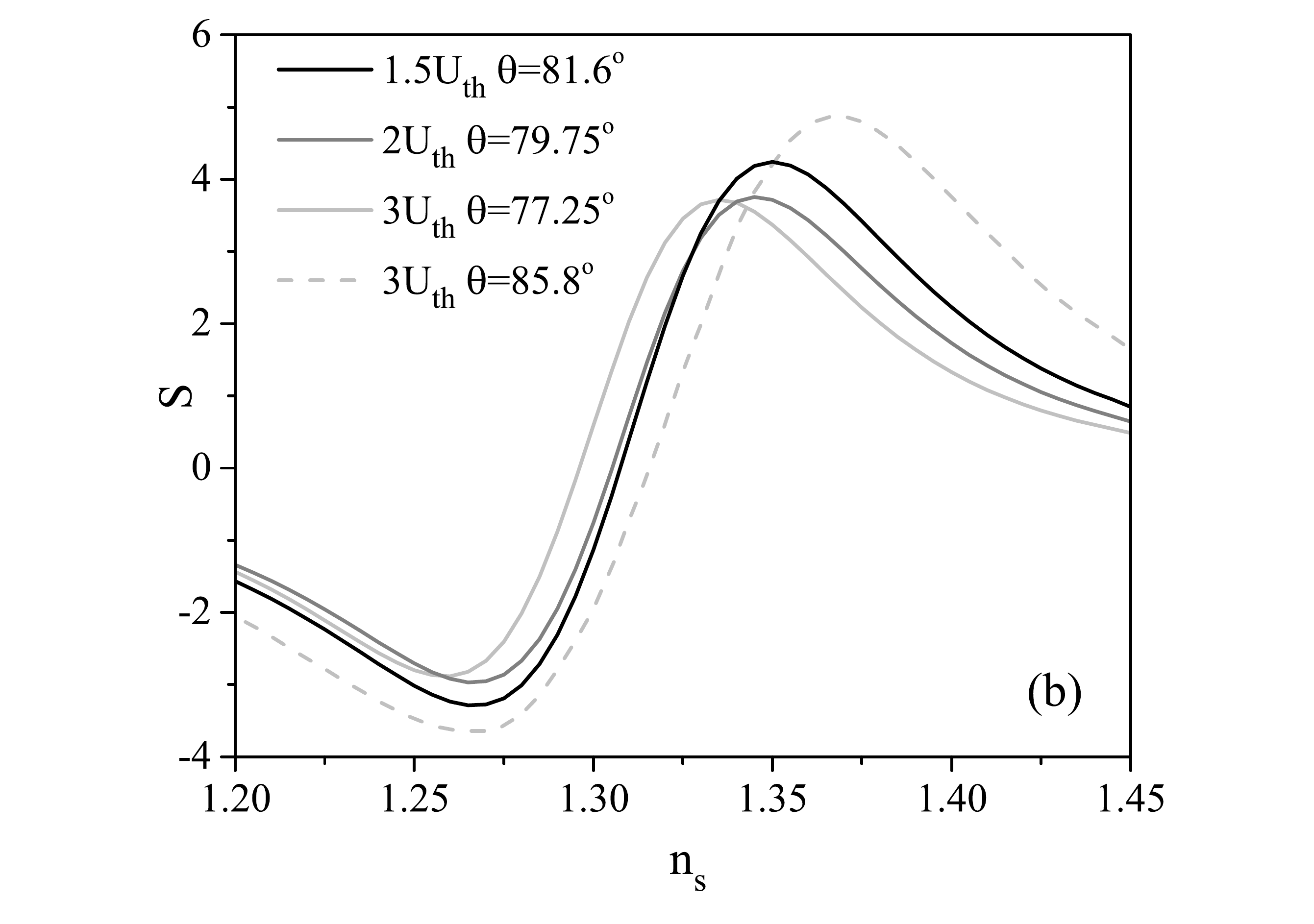}}
\caption{Sensitivity $S$ versus the analyte refractive index $n_{\text{s}}$ for 5CB~(a) and E44~(b) at voltages 1.5, 2 and 3$U_{\text{th}}$, $n_1=1.57$.}
\label{fig-8}
\end{figure}

We calculated the reflectance versus the incident angle at voltages 1.5, 2 and 3$U_{\text{th}}$ for $n_1$ from 1.53 to 1.63. When the $n_1$ value is higher than 1.6, the reflectance dips are too narrow. For a detailed investigation, the prism refractive index $n_1=1.57$ was chosen, because it shows a better sensitivity and there exists a real optical glass with close refractive index \cite{27}. The reflectance at voltages 1.5, 2 and $3U_{\text{th}}$ [figure~\ref{fig-3}~(a)] is shown in figure~\ref{fig-7}.
At the voltage $1.5U_{\text{th}}$, the reflectance has narrow dips similar to the planar case. When the voltage increases, the dips become wider. Similar behaviour can be observed when the prism refractive index changes at the fixed voltage: the dips become wider when $n_1$ decreases. For curves in figure~\ref{fig-7},  more than only last dips are suitable for measurements. Figure~\ref{fig-8} shows sensitivity versus the analyte refractive index at voltages 1.5, 2 and 3$U_{\text{th}}$ for 5CB (a) and E44 (b). For 5CB, the maximum sensitivities are as follows: $S_{\text{max}}=4.66$ at $n_{\text{s}}=1.355$ ($\theta=84.1^\circ$, $U=1.5U_{\text{th}}$) and $S_{\text{max}}=4.8$ at $n_{\text{s}}=1.37$ ($\theta=85.9^\circ$, $U=2U_{\text{th}}$).
For E44 the maximum sensitivities are as follows: $S_{\text{max}}=4.24$ at $n_{\text{s}}=1.35$ ($\theta=81.6^\circ$, $U=1.5U_{\text{th}}$) and $S_{\text{max}}=4.87$ at $n_{\text{s}}=1.37$ ($\theta=85.8^\circ$, $U=3U_{\text{th}}$).

\begin{figure}[!t]
\centerline{\includegraphics[width=0.5\textwidth]{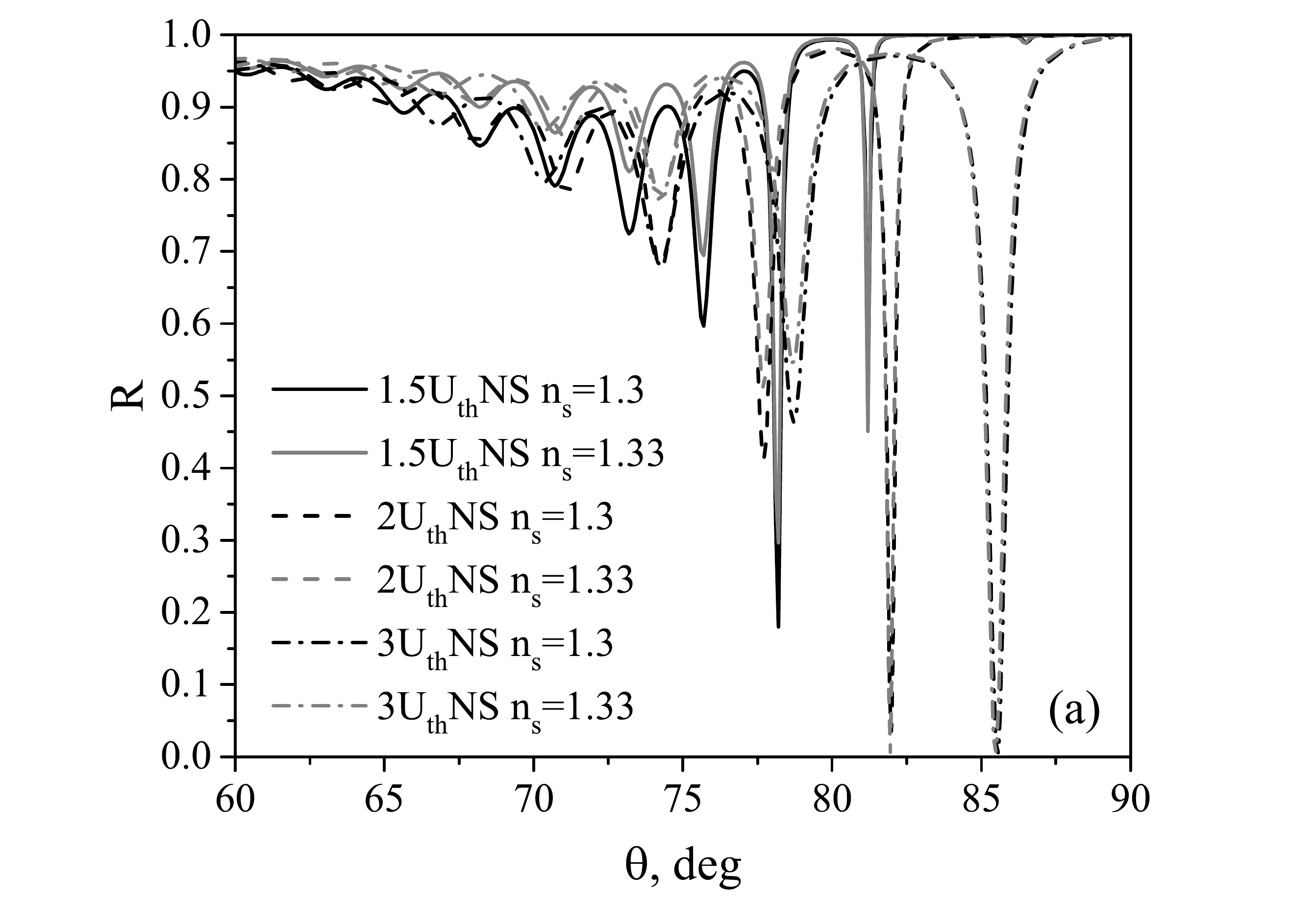}\includegraphics[width=0.5\textwidth]{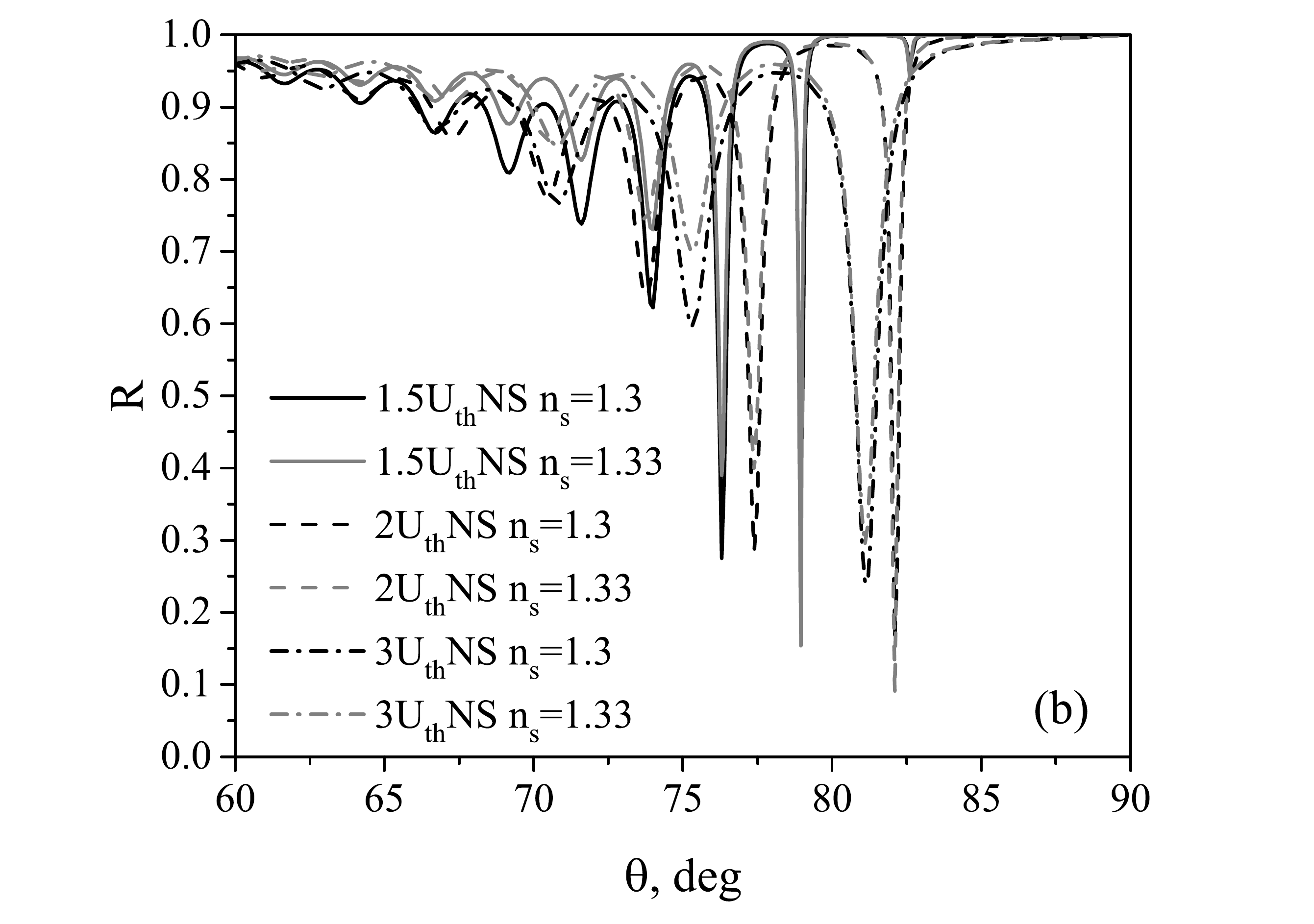}}
\caption{Reflectance $R$ versus the light incident angle  $\theta$ for 5CB (a) and E44 (b) at voltages 1.5, 2 and 3$U_{\text{th}}NS$, $n_1=1.57$, $n_{\text{s}}=1.3,1.33$.}
\label{fig-9}
\end{figure}

\begin{figure}[!t]
\centerline{\includegraphics[width=0.5\textwidth]{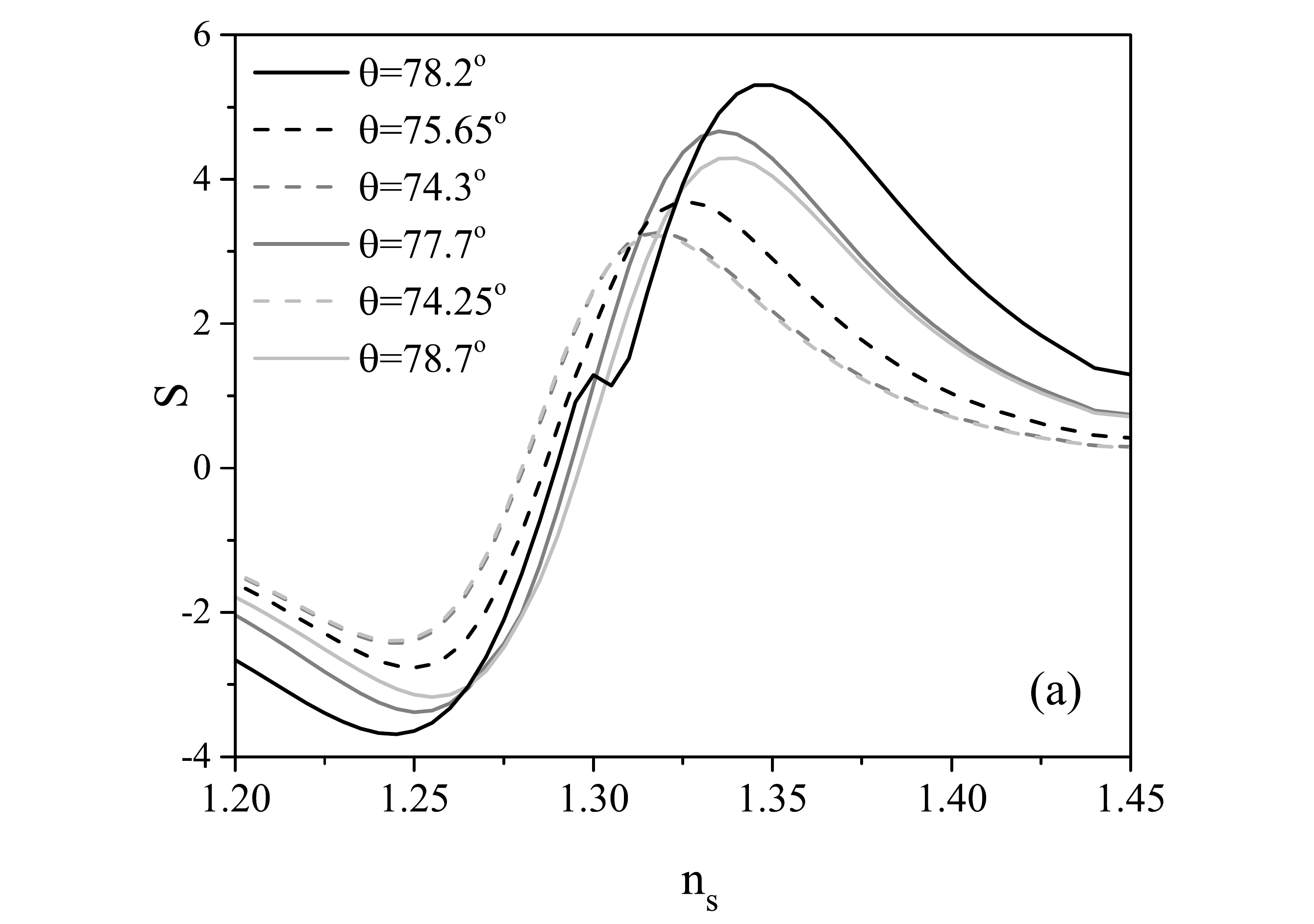}\includegraphics[width=0.5\textwidth]{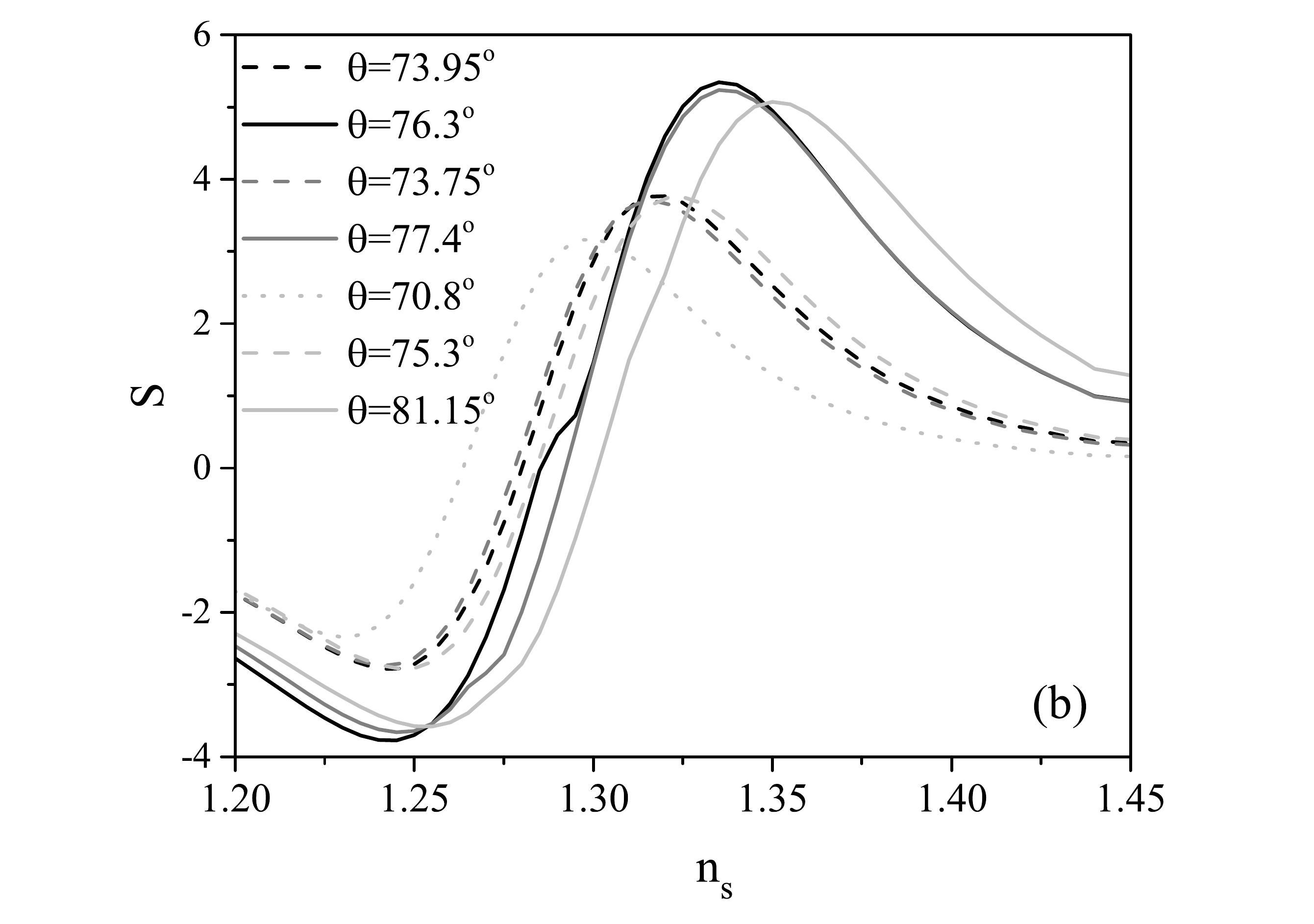}}
\caption{Sensitivity $S$ versus the analyte refractive index $n_{\text{s}}$ for 5CB (a) and E44 (b) at voltages 1.5, 2 and 3$U_{\text{th}}NS$ (black, gray and light gray, respectively), $n_1=1.57$.}
\label{fig-10}
\end{figure}

The reflectance for non-symmetrical cell at voltages 1.5, 2 and 3$U_{\text{th}}NS$ is shown in figure~\ref{fig-9}. The curves are similar to figure~\ref{fig-7} but $R(\theta)$ dips are deeper than in the symmetric case. Sensitivity versus the analyte refractive index at voltages 1.5, 2 and 3$U_{\text{th}}NS$ is shown in figure~\ref{fig-10} for 5CB (a) and E44 (b). For 5CB, maximum sensitivities are as follows: $S_{\text{max}}=5.3$ at $n_{\text{s}}=1.35$ ($\theta=78.2^\circ$, $U=1.5U_{\text{th}}NS$) and $S_{\text{max}}=4.67$ at $n_{\text{s}}=1.335$ ($\theta=77.7^\circ$, $U=2U_{\text{th}}NS$). For E44, maximum sensitivities are as follows: $S_{\text{max}}=5.34$ at $n_{\text{s}}=1.335$ ($\theta=76.3^\circ$, $U=1.5U_{\text{th}}NS$), $S_{\text{max}}=5.23$ at $n_{\text{s}}=1.335$ ($\theta=73.75^\circ$, $U=2U_{\text{th}}NS$) and $S_{\text{max}}=5.1$ at $n_{\text{s}}=1.35$ ($\theta=81.15^\circ$, $U=3U_{\text{th}}NS$). Thus, the highest sensitivity values were obtained for non-symmetrical cell at $n_1=1.57$.

In spite of the case $n_1=1.51$, in the case $n_1=1.57$, the maximum sensitivity position shifts depending on the voltage and the selected dip (for the selected angle of incidence). By varying the voltage, one can choose a better sensitivity depending on the analyte. In the sensor considered in \cite{13}, to measure the analyte refractive index higher than 1.33, one should take the prism with a higher refractive index, for example 1.78 for $n_{\text{s}}=1.45$. Since the effects of the LC director reorientation and the prism refractive index variation are similar, it is appropriate to use LCs instead of replacing the prism. Although we did not obtain a sensitivity increase at $n_{\text{s}}=1.45$, further optimization of system parameters and LC geometry can enhance the situation.

\section{Conclusions}
In this study we theoretically investigated the nanorod-mediated surface plasmon sensor with an inhomogeneous LC layer. Using the Berreman method, the light reflectance from multilayer system was calculated as a function of the incident angle for 5CB and E44 LCs at two types of boundary conditions. The reflectance has dips that could be explained by mixing of plasmon mode and half-leaky modes that propagate in the LC layer. Calculations were carried out at two values of the prism refractive index. In the case when the prism refractive index equals 1.51, for both boundary conditions there is a dip whose position is constant for the chosen LC configuration and its depth depends on the analyte refractive index. 
The analyte refractive index can be found from the reflectance value at the incident angle that corresponds to this reflectance dip. The reflectance at different applied voltages shows that the LC director reorientation allows one to control the dip position. Therefore, by changing the applied voltage, one can control the position of the reflective dips and arrange the measurements in the most convenient way.
In the case of the non-symmetrical boundary conditions, a stronger reorientation effect was not obtained. The sensor sensitivity to the analyte refractive index change was calculated as the ratio between the reflectance change and the analyte refractive index change. For E44, reorientation causes the sensitivity increase, but for 5CB, the sensitivity of homogeneous LC is higher. For both LCs, reorientation does not have a significant impact on the $n_s$ value at which the sensitivity has a maximum. 
The case of the value of the prism refractive index being between ordinary and extraordinary refractive indices of the LCs was also studied. This case presents an interest because the LC reorientation can change the ratio between the refractive indices of the layers and the effect of the LC director reorientation should be pronounced stronger. At $n_1=1.57$ without voltage, there is a region in the reflectance spectrum with constant reflectance $R=1$ that disappears when the voltage higher than the threshold is applied. In this case, the number of dips suitable for measurement increases, and when the voltage increases, the dips become wider. A decrease of the prism refractive index with a constant voltage has the same effect on the reflectance features. 
The improvement in the proposed measurement scheme is due to the fact that the voltage change and the resulting reorientation of the LC is easier than replacing the prism.
When $n_1=1.57$, the LC reorientation allows one to shift the maximum sensitivity. Such a sensor can be used for measuring in two steps. The first step is to measure the region of the analyte refractive index, the second one is to choose the voltage and make a more accurate measurement. A better sensitivity was obtained at $n_1 = 1.57$ in the case of non-symmetrical boundary conditions. Our results can be used for designing tunable sensors.


%
%

\ukrainianpart
\title{Числове моделювання поширення світла в сенсорі на основі поверхневого плазмонного резонансу з рідким кристалом}
\author{Є.С. Ярмощук, В.І. Задорожний, В.Ю. Решетняк}
\address{Київський національний університет імені Тараса Шевченка, Фізичний факультет, \\ просп. Академіка Глушкова, 2, 03022 Київ, Україна}
\makeukrtitle
\begin{abstract}
\tolerance=3000%
Теоретично досліджено п'ятишаровий плазмонний сенсор з шаром наночастинок та неоднорідним шаром рідкого кристалу. Розраховано коефіцієнт відбивання як функцію кута падіння при різних значеннях показника заломлення аналіту та прикладеної до рідкого кристалу (РК) напруги. Змінюючи орієнтацію директора рідкого кристалу, можна контролювати положення мінімумів кривих відбивання і вибрати той, що найбільш чутливий до показника заломлення аналіту. При вибраному куті падіння показник заломлення аналіту може бути визначений по значенню коефіцієнта відбивання. Вплив переорієнтації директора сильніший, коли значення показника заломлення призми є між звичайним і незвичайним показниками заломлення РК, в цьому випадку збільшення напруги і зменшення показника заломлення призми мають подібний вплив на форму кривої відбивання.
\keywords рідкий кристал, поверхневий плазмонний резонанс, сенсор, наночастинки, металева пориста плівка

\end{abstract}

\end{document}